\documentclass[aps,prl,superscriptaddress,showpacs,twocolumn]{revtex4}
\usepackage{amsfonts}
\usepackage{amsmath}
\usepackage{amssymb}
\usepackage{graphicx}

\begin{document}

\title{Orientation dependence of the optical
spectra in graphene at high frequencies }
\author{Chao Zhang$^{a,}$}
\affiliation{School of Engineering Physics, University of Wollongong, New South Wales
2552, Australia}
\author{Lei Chen}
\affiliation{School of Physics, Peking University, Beijing 100871, China}
\author{Zhongshui Ma$^{b,}$}
\affiliation{School of Physics, Peking University, Beijing 100871, China}
\affiliation{School of Engineering Physics, University of Wollongong, New South Wales
2552, Australia}

\begin{abstract}
On the basis of the Kubo formula we evaluated the optical
conductivity of a graphene sheet. The full behavior of frequency
as well as temperature dependence of the optical conductivity is
presented. We show that the anisotropy of conductivity can be
significantly enhanced at high frequencies. The photon absorption
depends on the field polarization direction. At the frequency
comparable to the maximum separation of upper and lower bands the
photon-induced conduction of electrons is strongly suppressed if
the polarization of field is along the zigzag direction. The
corresponding optical conductivity is several orders of magnitude
weaker than that when the light is polarizing along the armchair
direction. We propose that the property of orientation selection
of absorption in the graphene can be used as a basis for a
high-frequency partial polarizer.
\end{abstract}

\maketitle

\medskip Because it is recognized as a new class of materials\cite{novo1} in
the carbon family, graphene has recently attracted intensive
interests as it possesses remarkable electronic properties, such
as the anomalous integer quantum Hall effect\cite{novo2,zhang1},
the finite conductivity at zero charge-carrier
concentration\cite{novo2}, the strong suppression of weak
localization\cite{suzuura,morozov,khveshchenko,mccann}, etc. These
properties promise building blocks for the technological
applications in molecular electronic and optoelectronic devices.
In graphene, the conduction and valance bands touch upon each
other at isolated points in the Brillouin zone ($K$ and $K^{\prime
}$). Undoped graphene is a gapless semiconductor, or a semimetal,
with vanishing density of states at the Fermi level. Low-energy
electronic states of graphene with a linear dispersion at the
corner of Brillouin zone are described by the "relativistic"
massless Dirac equation. This relativistic kinematical description
of graphene is confirmed in quantum Hall studies\cite{zheng} and
gives us theoretical insight into exotic transport\cite{peres},
magnetic correlation\cite{fujita,vozmediano}, and
dielectric\cite{sarma} properties observed in this material. On
the theoretical side, most of works in studies of graphene is
based on "massless Dirac theory" in the vicinity of the $K$ point
in the Brillouin zone, where all physical quantities remain
isotropic. The purpose of this letter is to reveal transport
properties properties beyond the regime of linear energy
dispersion. Because the carrier concentration in graphene can be
varied over a wide range by doping and by the electric field
effect\cite{novo1}, this makes high-energy excitations in graphene
important and interesting. We shall show that the current response
is highly isotropic away from the $K$-point and the high frequency
spectra can reveal a variety of rich physics and anomalous
phenomena that electron and holes possess in graphene.

Besides its transport properties, the gapless energy spectrum of electrons
and holes in graphene can lead to specific features of optical and plasma
properties\cite{ando,gusynin}. It is anticipated that the optical spectrum
induced by elementary electronic excitation can be used to determine the
electronic properties of graphene. It is known that optical conductivity is
one of the central quantities to determine almost all optical properties of
an electron and/or hole system. For a case where the optical transition is
induced mainly by a dielectric response of the carriers through the
carrier-carrier interaction, the optical conductivity can be obtained simply
from the Kubo formula in which the current-current correlation is mainly
caused by carrier interactions with a weak external light field\cite{mish}.

In the present letter we present one interesting property of
graphene at high frequencies. On the basis of current-current
correlation function we investigate the fast-electron optical
spectrum of graphene. We obtained the full frequency and
temperature dependence of the optical spectra for graphene with
varying Fermi energy. The optical spectrum show a cusp-like
maximum at $\hbar \omega _{c}=2\times \left( 2\hbar
v_{F}/\sqrt{3}a\right) $, where $v_{F}$ is Fermi velocity and $a$
($=\sqrt{3}b$ with $b$ the c-c bond length) is the lattice
constant. However, the absorption beyond $\omega _{c}$ is
negligible for photons polarizing along the zigzag direction. For
light polarizing along the armchair direction there is an
absorption continuum which jumps abruptly at 3$\hbar \omega _{c}$,
corresponding to the vertical transition at $\mathbf{k}=0$. Our
results demonstrate that graphene can be used as a partial
polarizer in the high frequency regime where the absorption along
the armchair direction is finite while the absorption along the
zigzag direction is basically zero. We analyze this conductivity
anisotropy in terms of the energy dependence of the interband
velocity operator.

\begin{figure}[h]
\centering \includegraphics[width=8.0cm]{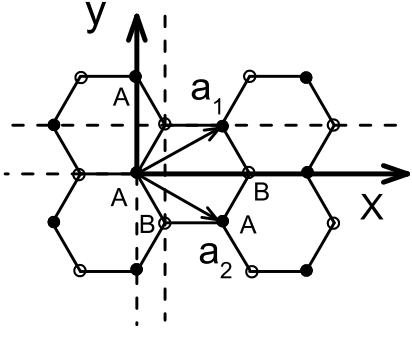}
\caption{Polarization direction in graphene, x-axis is the armchair
direction and y-axis is the zigzag direction.}
\label{fig1}
\end{figure}

A unit cell, whose wave vectors are given by $\mathbf{a}_{1}=a\left( \sqrt{3}%
/2,1/2\right) $ and $\mathbf{a}_{2}=a\left( \sqrt{3}/2,-1/2\right)
$, in a honeycomb lattice contains two atoms denoted as A and B.
In this letter the y-axis is considered to be along the line
connecting atoms of same sublattice while the x-axis along the
line connecting atoms of alternating sub-lattice, as shown in
Fig.1. We assume the graphene is intrinsic, i.e., free of impurity
doping. For the case of zero carrier density the two bands cross
the Fermi level (setting as zero) at $\mathbf{K}$ and
$\mathbf{K}^{\prime }$ points of the first Brillouin zone, whose
wave vectors are given by $\mathbf{K}=\left( 4\pi /3a\right)
\left( 1/\sqrt{3},1\right) $ and $\mathbf{K}^{\prime }=\left( 4\pi
/3a\right) \left( 3/2\sqrt{3},-1/2\right) $. In the presence of
applied electric field, the chemical potential can be moved to
either positive or negative regime. The Hamiltonian of graphene is
given as
\begin{equation}
H_{0}=\left(
\begin{array}{cc}
0 & h(\mathbf{k}) \\
h^{\ast }(\mathbf{k}) & 0%
\end{array}%
\right)
\end{equation}%
with $h(\mathbf{k})=-t\left[ 1+e^{i\mathbf{k}\cdot \mathbf{a}_{1}}+e^{i%
\mathbf{k}\cdot \mathbf{a}_{2}}\right] .$ The eigen functions and the eigen
values can be written as $\xi _{\mathbf{k},s}=\left( 1/\sqrt{2}\right) \left(
-s\left( t/\left\vert t\right\vert \right) e^{i\varphi _{\mathbf{k}%
}},1\right) ^{T}$ and $\epsilon _{\mathbf{k,}s}=s\epsilon
_{0}\sqrt{3+\alpha \left( \mathbf{k}\right) }$, where $s=\pm 1$,
$\tan \varphi =\tan ^{-1}\left(
\sin \mathbf{k}\cdot \mathbf{a}_{1}+\sin \mathbf{k}\cdot \mathbf{a}%
_{2}\right) /\left( 1+\cos \mathbf{k}\cdot \mathbf{a}_{1}+\cos \mathbf{k}%
\cdot \mathbf{a}_{2}\right) $, $\alpha \left( \mathbf{k}\right) =-2+4\cos
\sqrt{3}k_{x}a/2\cos k_{y}a/2+4\cos ^{2}k_{y}a/2$, and $\epsilon _{0}=2\hbar
v_{F}/\sqrt{3}a$. $v_{F}$ is the Fermi velocity, which relates to the
hopping parameter, i.e., $v_{F}=i\left( \sqrt{3}a/2\hbar \right) t$.

In the presence of a uniform time-dependent electric field $\mathbf{E=E}%
_{0}e^{-i\omega t}$, the Hamiltonian becomes $H=H_{0}+H^{\prime }$ with $%
H^{\prime }=\sum_{i=1,2}\left(
\begin{array}{cc}
0 & -it\frac{e}{\hbar c}e^{i\mathbf{k}\cdot \mathbf{a}_{i}} \\
it^{\ast }\frac{e}{\hbar c}e^{-i\mathbf{k}\cdot \mathbf{a}_{i}} & 0%
\end{array}%
\right) \mathbf{A}\cdot \mathbf{a}_{i},$where the vector potential is
written as $\mathbf{A}=(c/i\omega )\mathbf{E}$.
By using the Fermi field operator $\widehat{\Psi }\left( \mathbf{%
r}\right) =\left( 1/4\pi ^{2}\right)
\sum_{\mathbf{k},s}a_{\mathbf{k},s}\xi
_{\mathbf{k},s}e^{i\mathbf{k}\cdot \mathbf{r}}$, the second
quantized
Hamiltonian is obtained $H=\sum_{\mathbf{k},s}\epsilon _{\mathbf{k,}s}a_{%
\mathbf{k},s}^{\dag }a_{\mathbf{k},s}+\left( 1/c\right)
\mathbf{A}\cdot \mathbf{J}$.  The components of the current
density operator $\mathbf{J}$ can be written as,
\begin{widetext}
\begin{equation}
J_{x}=-2ev_{F}\sum_{\mathbf{k},s}s\frac{\sin \frac{\sqrt{3}}{2}k_{x}a\cos
\frac{1}{2}k_{y}a}{\sqrt{3+\alpha \left( \mathbf{k}\right) }}a_{\mathbf{k}%
,s}^{\dag }a_{\mathbf{k},s}+i2ev_{F}\sum_{\mathbf{k},s}s\frac{1+\cos \frac{%
\sqrt{3}}{2}k_{x}a\cos \frac{1}{2}k_{y}a+\cos k_{y}a}{\sqrt{3+\alpha \left(
\mathbf{k}\right) }}a_{\mathbf{k},s}^{\dag }a_{\mathbf{k},-s}
\end{equation}%
\end{widetext}and%
\begin{widetext}
\begin{equation}
J_{y}=-\frac{2ev_{F}}{\sqrt{3}}\sum_{\mathbf{k},s}s\frac{\cos \frac{\sqrt{3}%
}{2}k_{x}a\sin \frac{1}{2}k_{y}a+\sin k_{y}a}{\sqrt{3+\alpha \left( \mathbf{k%
}\right) }}a_{\mathbf{k},s}^{\dag }a_{\mathbf{k},s}-i\frac{2ev_{F}}{\sqrt{3}}%
\sum_{\mathbf{k},s}s\frac{\sin \frac{\sqrt{3}}{2}k_{x}a\sin \frac{1}{2}k_{y}a%
}{\sqrt{3+\alpha \left( \mathbf{k}\right) }}a_{\mathbf{k},s}^{\dag }a_{%
\mathbf{k},-s}.
\end{equation}%
\end{widetext}
From the above expression it is seen that for a graphene under an
optical field, two kinds of transitions (intra and interband)
contribute to the current response. The Kubo formula for the
dynamic conductivity is given as
\begin{equation}
\sigma _{\mu \nu }\left( \omega
\right) =\left( 1/\omega \right) \int_{0}^{\infty }dte^{i\omega
t}\left\langle \left[ J_{\mu }\left( t\right) ,J_{\nu }\left( 0\right) %
\right] \right\rangle.
\end{equation}
Based on the current operators given in Eqs. (2) and (3), it is
found that the intraband terms do not contribute to the
conductivity in graphene free of any disorders. The electron
conduction is solely due to the electron jumping between the bands
with the absorption of a photon. After some algebra, we found,
\begin{eqnarray}
\sigma _{xx} &&=\frac{e^{2}v_{F}^{2}}{\pi ^{2}\omega }\int d\mathbf{k}\frac{%
\left( 1+\cos \frac{\sqrt{3}}{2}k_{x}a\cos \frac{1}{2}k_{y}a+\cos
k_{y}a\right) ^{2}}{3+\alpha \left( \mathbf{k}\right) }  \notag \\
&&\left( \frac{f_{\mathbf{k},+}-f_{\mathbf{k},-}}{\hbar \omega +2\epsilon
_{0}\sqrt{3+\alpha \left( \mathbf{k}\right) }}-\frac{f_{\mathbf{k},+}-f_{%
\mathbf{k},-}}{\hbar \omega -2\epsilon _{0}\sqrt{3+\alpha \left( \mathbf{k}%
\right) }}\right)
\end{eqnarray}%
and
\begin{eqnarray}
\sigma _{yy} &&=\frac{e^{2}v_{F}^{2}}{3\pi ^{2}\omega }\int d\mathbf{k}\frac{%
\sin ^{2}\frac{\sqrt{3}}{2}k_{x}a\sin ^{2}\frac{1}{2}k_{y}a}{3+\alpha \left(
\mathbf{k}\right) }\left( f_{\mathbf{k},+}-f_{\mathbf{k},-}\right)  \notag \\
&&\left[ \frac{1}{\hbar \omega +2\epsilon _{0}\sqrt{3+\alpha \left( \mathbf{k%
}\right) }}-\frac{1}{\hbar \omega -2\epsilon _{0}\sqrt{3+\alpha \left(
\mathbf{k}\right) }}\right]
\end{eqnarray}%
with $\omega =\omega +i0_{+}$, where $f_{\mathbf{k},s}$ is Dirac-Fermi
distribution function. In addition, $\sigma _{xy}=\sigma _{yx}=0$ because of
symmetry.

\begin{figure}[h]
\centering \includegraphics[width=9cm]{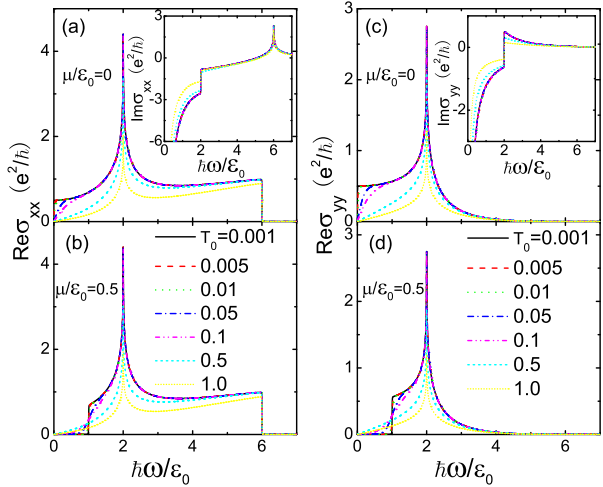} \caption{(a) and
(c): The real part of the optical conductivity
$\protect\sigma_{xx}$ and $\protect\sigma_{yy}$ vs frequency at
various temperatures for the Fermi energy lies on $K$-point.
Inset: the imaginary part of the optical conductivity
$\protect\sigma_{xx}$ and $\protect\sigma_{yy}$ vs frequency. (b)
and (d): Same as (a) and (c) but with the Fermi energy lies in the
conduction band.} \label{fig2}
\end{figure}

From the forms of dynamic conductivity we identify immediately our
first result that the optical absorption for photons polarizing
along x-direction is quite different from that when photons are
polarizing along y-direction. To see this polarization dependence
of the conductivity we calculate the real and imaginary parts of
dynamic conductivity numerically. With an energy unit $ \epsilon
_{0}$ the temperature is scaled as $T_{0}=k_{B}T/\epsilon _{0}$.
The frequency dependence of real
part of the optical conductivity $\mathrm{Re}\sigma _{xx}$ and $\mathrm{Re}%
\sigma _{yy}$ are shown in Figs. 2, with two different chemical
potentials. It is found that the photon absorption starts from
$\omega =0$ and increases with $\omega $ for a graphene with $\mu
=0$ (Figs. 2(a) and 2(c)) at low temperature, which is consistent
with the metallic absorption characteristics. However, for a
graphene with nonzero $\mu $ (see in Figs. 2(b) and 2(d)) there is
a threshold frequency $\omega _{th}=2\mu /\hbar $ and the
absorption is almost negligible when the photon frequency is below
$\omega_{th}$. The reason is that the allowed transitions only
involve vertical
processes, $\hbar \omega =\epsilon _{\mathbf{k,}s}-\epsilon _{\mathbf{k,}%
s^{\prime }}$. As temperature increases, the phase space for
electronic transitions for $\omega <2\mu $ increases and finite
absorption is observed. Both $\mathrm{Re}\sigma _{xx}$ and
$\mathrm{Re}\sigma _{yy}$ exhibit a sharp absorption maximum at
$\omega =2\epsilon _{0}/\hbar $. This corresponds to vertical
transitions at $k_{y}a=\pi $ in the energy contour of the
reciprocal lattice. The joint density of state (JDOS) for
electron-hole excitations reaches the cusp-like maximum at this
energy due to the maximum $k_x$-degeneracy at $k_{y}a=\pi $. As a
result the absorption is sharply peaked with a cusp shape. When
the frequency $\omega $ exceeds $2\epsilon _{0}/\hbar $ it is
found that $\mathrm{Re}\sigma _{xx} $ and $\mathrm{Re}\sigma
_{yy}$ have very different frequency dependence. The $\omega
$-dependence of $\mathrm{Re}\sigma _{xx}$ increases slowly with
$\omega $ at high $\omega $ and jumps abruptly to zero at $\omega
=6\epsilon _{0}/\hbar $, a feature due to the interplay of the
increase interband velocity component $v_x$ and decreasing DOS.
The jumping point corresponds to the maximum energy gap in energy
contour. In contrast to the $\mathrm{Re}\sigma _{xx}$ there is no
jumping in the $\mathrm{Re}\sigma _{yy}$ and it tends to vanish
rapidly for $\omega >2\epsilon _{0}/\hbar $. The reason for this
striking difference between the $\mathrm{Re}\sigma _{xx}$ and the
$\mathrm{Re}\sigma _{yy}$ at high frequency lies in the energy
dependence of the interband velocity operator $\mathbf{v}_{s,-s}$
(referring to the second term of Eqs. 2 and
3). For the $x$-component of current $j_{x}$, the corresponding velocity $%
v_{s,-s}^{x}$ increases as energy increases and reaches its maximum which
coincides with the maximum separation of energy between conduction and
valance bands at $\mathbf{k}=0$. Differing from $v_{s,-s}^{x}$ the $%
v_{s,-s}^{y}$ for $j_{y}$ decreases rapidly with energy at high frequency.
As a result, although the absorbing y-polarized photon at high frequency is
energetically possible, the excited electrons are not allowed to gain any
velocity along y-direction.

The cusp-like maximum in the optical spectral is unique to
honeycomb lattice. The derivative at cups is discontinuous.
According to the Kramers-Kronig relations, a discontinuous
derivative in the real part requires that the imaginary part has a
step-like behaviors in the imaginary part, and vice versa. This
has been confirmed by the inset in Figs. 2(a) and 2(c).

\begin{figure}[h]
\centering \includegraphics[width=9.0cm]{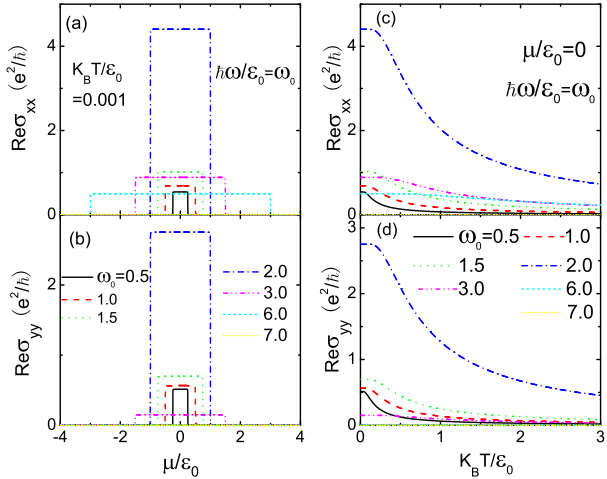} \caption{(a)
and (b): The real part of the optical conductivity
$\protect\sigma_{xx}$ and $\protect\sigma_{yy}$ vs chemical
potential at various frequencies. (c) and (d): The real part of
the optical conductivity $\protect\sigma _{xx}$ and
$\protect\sigma _{yy}$ vs temperature at various frequencies.}
\label{fig4}
\end{figure}

With the field effect the chemical potential can be tuned upper or
lower than $K$-point. The dependence
of the optical spectra on the chemical potential is shown
in Fig. 3a and 3b. At low
temperature, the $\mu $-dependencies of both $\mathrm{Re}\sigma _{xx}$ and $%
\mathrm{Re}\sigma _{yy}$ are step-like functions. For a given
$\omega $, the spectrum is nonzero within the regime of $-\hbar
\omega /2<\mu <\hbar \omega /2$. The height of the rectangular
spectra reflects the frequency dependence. At low temperature, the
conductivity takes on a value of order $\pi e^{2}/h$, which can be
viewed as an frequency-dependent analog of minimum
conductivity\cite{novo2}.

\begin{figure}[h]
\centering \includegraphics[width=9cm]{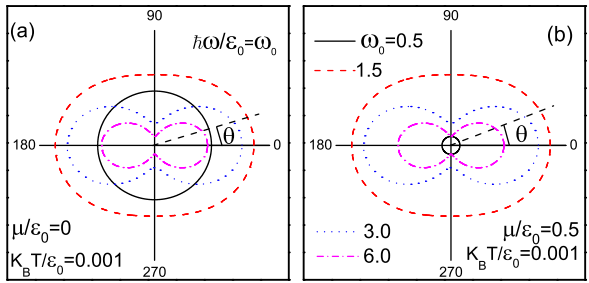} \caption{The
angular dependence of the absorption coefficient,
$\mathrm{Re}\sigma_{l}(\theta)$.} \label{fig5}
\end{figure}

\begin{figure}[h]
\centering \includegraphics[width=9cm]{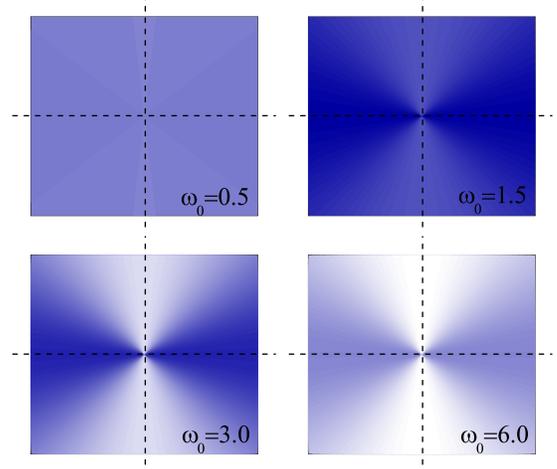} \caption{The
angular dependence of the transmitted field. The lightness
represents the absolute intensity.} \label{fig6}
\end{figure}

In addition the optical conductivity strongly depends on
temperature. In Fig. 3c and 3d, we show the temperature dependence
of the optical spectra. At fixed frequency, it is found that the
optical spectra shows a narrow plateau at low temperature then
decreases rapidly with temperature. The number of carrier in the
valance band at a given transition energy, $\epsilon $, is
proportional to $\exp (-\epsilon /k_{B}T)$. Therefore the the
optical spectra decay exponentially with the increase of
temperature. The width of the initial plateau is broaden with
increasing of $\omega $, which indicates that the low-$T$ thermal
smearing effect is less for high energy states.

The electromagnetic absorption is given by the real part of the
longitudinal conductivity along arbitrary direction. The
polarization dependent absorption strength can be best viewed in
the contour diagram shown in fig.4. The anisotropy is slightly
stronger for $\mu \neq 0$ The angular variation of the absorption
leads to a transmitted field whose intensity changes with the
polarization direction. The relative field intensity of the
transmitted field through a graphene is shown in Fig.5. The higher
the frequency is, the stronger the anisotropy in the transmitted
field.

In conclusion, we have studied the optical conductivity of graphene and
shown its orientation dependence in frequency for photons polarizing along
the armchair direction or along the zigzag direction. We show evidently that
the absorption vanishes rapidly for frequencies above $2\epsilon _{0}$, in
comparison of that along armchair direction, if the photons are polarized
along zigzag direction. As a consequence, the orientation selection of
excited electron transport indicates that the graphene behaves like a
partial polarizer at high frequencies. So the characteristics of property
presented in the letter is suggested to be measured via optical absorption
experiments.

This work is supported in part by the Australian Research Council,
NNSF-China under Grant No. 10674004, and NBRP-China under Grant No.
2006CB921803.

\noindent{$^{a)}$Electronic mail: czhang@uow.edu.au}

\noindent{$^{b)}$Electronic mail: mazs@phy.pku.edu.cn}

\end{document}